\documentstyle[12pt]{article}
\begin{document}

\begin{flushright}
hep-ph/9710425\\ October 1997
\end{flushright}

\vspace{9mm}

\begin{center}

{\Large \bf Using muons to probe for new physics\footnote{Plenary 
talk  at the 5th
International Conference {\em Beyond the Standard Model},
Balholm, Norway, Apr.~29th -  May 4th, 1997.}}

\vspace{9mm}

{\large\bf Andrzej Czarnecki} 

\vspace{2mm}

{\it Physics Department\\
Brookhaven National Laboratory\\  Upton, New York 11973}
\end{center}

\vspace{19mm}

\begin{abstract}
Searches for new physics using muons are reviewed. Particular
attention is paid to muon number non-conserving processes, like the
decay $\mu\to e\gamma$ and muon--electron conversion in muonic atoms.
Also, experimental determinations and theoretical predictions for the
muon anomalous magnetic moment are reviewed.
\end{abstract}

Muons were first observed in cosmic ray detectors in 1936
\cite{Anderson:36,Street:37}.  However, for several following years
they were being confused with pions, strongly interacting mesons
predicted by Yukawa.  It was not until 1947 that the identity of the
muon as a weekly interacting particle was demonstrated
\cite{Conversi:47}.  Evidence for the existence of two kinds of
charged particles with similar masses was provided by a cosmic ray
experiment by Powell and collaborators~\cite{Powell:47} (for the early
history of the muon physics see \cite{WuHughes,Nishijima63}).

Muons have proved to be extremely useful both in fundamental studies
and as a tool in applied science.  They are relatively long-living
(lifetime 2.2 $\mu s$) and can be produced in abundance so that
intense muon beams can be obtained.  In the latter aspect we expect
future improvements by several orders of magnitude which may lead to
construction of muon colliders, a new generation of particle
accelerators.

After 50 years of rich and varied muon physics programs it is
appropriate to review what has been achieved so far and what are the
future prospects.  This talk is but a modest contribution towards this
goal. I will focus on two aspects of muon physics, the anomalous
magnetic moment measurements and muon number non-conservation searches.

\section{Muon number non-conservation}
Lepton number conservation is an empirical law incorporated into the
standard model.  Most ideas about a more fundamental theory predict
violations either of individual (electron, muon, tau) numbers or of
the total lepton number.  Reactions in which muon number is violated,
such as $\mu\to e\gamma$ or muon--electron conversion in the field of
a nucleus, are a sensitive probe of ``new physics'' scenarios.  Table
\ref{tab:muonviol} lists examples of experimental tests of muon number
conservation in reactions in which a muon is present in the initial
state.  Other experimental searches are performed in meson decays,
such as $K_L\to \mu e$, $K_L\to \pi^0\mu e$, $K_L\to \mu^\pm\mu^\pm
e^\mp e^\mp$, $K^+\to \pi^+ \mu^+ e^-$, and $\pi^0\to \mu e$
\cite{PDG}.  Exotic $Z$ boson decays such as $Z\to \mu e$ have been
searched for at LEP; bounds obtained there are several orders of
magnitude weaker than in low energy experiments.  However, some
couplings like $Z\mu^\mp \tau^\pm$ can be better constrained by the LEP
experiments.

\begin{table}[htbp]
  \caption{Experimental searches for muon number non-conservation}
  \label{tab:muonviol}
  \begin{center}
    \leavevmode
    \begin{tabular}[b]{lccc}
\hline\hline
Reaction & Current Bound & Ongoing efforts & Proposals \\ \hline
$R(\mu^-N\to e^-N)$ &  $< 7 \times 10^{-13}$ 
                    & $\sim 2\times 10^{-14}$
                    & $10^{-16}$ \\
                 & SINDRUM II (Ti) & SINDRUM II (Au, Ti) & MECO (BNL)
                 \\[1mm]  
$B(\mu^+ \to e^+e^-e^+)$ & $<1\times 10^{-12}$ & --- & --- \\[1mm] 
     & SINDRUM \protect\cite{Bellgardt:88} & --- & --- \\[1mm] 
$B(\mu^+ \to e^+ \gamma)$ & $<4.2\times 10^{-11}$ & $\sim 7\times 10^{-13}$ 
                             & $10^{-14}$  \\
 &MEGA: 1993 data  & MEGA: optimistic goal & PSI \\[1mm] 
$R(\mu^+ e^- \to \mu^- e^+)$ & $<7.9\times 10^{-9}$ & $10^{-11}$ & \\
 & PSI \protect\cite{Jungmann95} & PSI \protect\cite{Jungmann89} &
\\[1mm] 
$B(\mu \to e\gamma\gamma)$ &  $<7.2\times 10^{-11}$ &    & 
\\[1mm] 
$R(\mu^-N_1\to e^+N_2)$ &  $< 8.9 \times 10^{-11}$ 
                    & $\sim 10^{-12}$
                    &           \\
 & SINDRUM II (Ti/Ca) & SINDRUM II \protect\cite{dohmen:1993} &
                 \\[1mm]  
\hline\hline    \end{tabular}
  \end{center}
\end{table}

\subsection{The decay \boldmath $\mu\to e\gamma$}
In the first years of muon physics the decay $\mu\to e\gamma$ was
considered as a candidate for the dominant decay channel.  However, it
has soon been demonstrated that the neutral component of the final
state in the normal muon decay is not a photon.  At the end of the
1950s a hypothesis of an intermediate vector boson (IVB) was proposed.
It was then thought that $\mu\to e\gamma$ might arise as a loop effect
of an IVB.  Lack of experimental evidence for such a decay led to the
hypothesis of two neutrinos carrying electron and muon flavor quantum
numbers, respectively.  This was confirmed in a famous experiment in
Brookhaven in 1962.  One can say that the experimental bounds on
$\mu\to e\gamma$ led to the concept of families, a cornerstone of the
standard model.

The experimental situation in $\mu\to e\gamma$ searches until the end
of 1972, as well as earlier references, can be found in
\cite{Frenkel72}.  More recent summary has been given in
\cite{Schaff93}.  Currently the best upper bound on the branching
ratio BR$(\mu\to e\gamma)<4.2\times 10^{-11}$ (at 90\% C.L.) comes
 from the experiment MEGA \cite{Mischke97}, based on data collected up
to 1993.  After more recent MEGA data have been analyzed, this bound
will improve.

The search for $\mu^+\to e^+\gamma$ in the contemporary experiments is
based on looking for simultaneous, collinear, back-to-back photon and
positron with energies equal half the muon mass, 52.8 MeV.  There are
two important sources of background.  A radiative decay of the muon,
$\mu^+\to e^+\nu_e\bar\nu_\mu\gamma$, has the same signature if the
neutrinos carry away little energy; this is the so-called physics
background.  In experiments with high rates of muons there may occur
also accidental background, with an energetic electron coming from a
normal decay of one muon, and a photon from a radiative decay of
another one.  Recently, a new idea has been put forward to reduce both
backgrounds \cite{Kuno:1997zw,Kuno:1996kv}, based on a careful study
of angular distributions of photons and electrons in normal and
radiative decays of polarized muons.  In the decay $\mu^+\to
e^+_{R,L}\gamma$, $e^+_R$ has an angular distribution
$1-P\cos\theta_e$ ($P$ is the degree of muon polarization and
$\theta_e$ is the angle between the direction of the muon spin and the
positron track), while for $e^+_L$ it is $1+P\cos\theta_e$.  On the
other hand, the highest energy positrons in the normal decay $\mu^+\to
e^+\nu_e\bar\nu_\mu$, as well as the highest energy photons in the
radiative decay, follow the distribution law
$1+P\cos\theta_{e,\gamma}$.

By looking for the positrons emitted in the direction antiparallel to
the muon spin one suppresses both the physics and the accidental
backgrounds for $\mu^+\to e^+_R\gamma$.  On the other hand, by
searching for photons emitted in that direction one can suppress the
accidental background for $\mu^+\to e^+_L\gamma$.

If polarized muons are used, it seems feasible to search for $\mu^+\to
e^+\gamma$ with the sensitivity of about $10^{-14}$.  At this level
one can perform an interesting test of supersymmetric grand
unification models \cite{Barbieri:1994pv,Barbieri:1995tw,Hall:1986dx}.
In addition to suppressing the backgrounds, experiments with polarized
muons could reveal the chirality of the produced electron, which could
help in determining the underlying mechanism for this exotic decay.  A
proposal is currently under preparation for a new $\mu^+\to
e^+\gamma$ search at Paul Scherrer Institute (PSI).

\subsection{Conversion of muons into electrons on nuclei}
\subsubsection{Muon--electron conversion: theory}
Theory of a muon conversion into electrons in the field of a nucleus
was first studied by Weinberg and Feinberg \cite{wf59}.  They focused
on an electromagnetic mechanism of transfering the energy yield to
the nucleus.  The structure of this electromagnetic interaction is
richer than for the $\mu\to e\gamma$ decay since the photon need not
be on mass shell.  Therefore, it is possible that the conversion can
occur even if $\mu\to e\gamma$ is exactly forbidden for some reason.
For example, the matrix element for the conversion contains monopole
terms which do not contribute to the decay $\mu\to e\gamma$; this is
because the longitudinal polarization states are possible only for
virtual photons.  In addition, there may be other mechanisms of the
conversion apart from the photon mediated processes.  This makes the
conversion on nuclei a particularly interesting process to study.

The early theoretical studies of the muon conversion into electrons on
nuclei  performed in \cite{wf59,Cabibbo59,rosen60} 
were valid mainly for conversion on light nuclei. 
 The
developments before the year 1978 have been summarized in
\cite{MarcianoOrbis}.  
In heavier atoms new effects become important: relativistic components
of the muon wave function, Coulomb distortion of the outgoing
electron, and the finite nuclear size.  These were addressed in
\cite{Shanker79}.  Nuclear effects were also analyzed, albeit in a
non-relativistic approximation, in \cite{Chiang93} and, more recently,
in \cite{Kosmas:1997a,Kosmas:1997b}.  

 There are two aspects of theoretical nature in describing the
muon--electron conversion, characterized by different distance scales.
The short distance effects which are responsible for the muon number
violation may be caused only by some as yet unknown ``new physics''.
However, the rate of the transition depends also on the long distance
atomic physics of the muonic atom.

The first group of problems has been studied in many extensions of the
standard model (for a review see
\cite{MarcianoOrbis,Kosmas:1994pt,Vergados:1986pq}).  In particular,
in ref.~\cite{Marciano77PRL} the rate of the coherent conversion $\mu
^-N \to e^-N $ was calculated in a variety of gauge models.  It was
pointed out in that paper that in a large class of models the
conversion can be much more probable than the decay $\mu\to e\gamma$.
This is because of the logarithmic enhancement of the form factors
leading to the conversion but absent in the decay rate.  Such
logarithmic effects were also recently discussed in
\cite{Santamaria97}.

The atomic physics aspects were studied in \cite
{Shanker79,Chiang93,Kosmas:1997a}.  Probably the most complete study
to date is given in \cite {Shanker79}, where relativistic wave
functions, Coulomb distortion, and finite nuclear size effects are
taken into account in the analysis of the coherent conversion.  In
many other calculations various aspects of the coherent conversion
have been considered, but none of those calculations covered all
potentially important effects to the extent the paper \cite
{Shanker79} did.  For example, in \cite{Chiang93,Kosmas:1997a} the
nuclear effects were analyzed but only in a nonrelativistic
approximation.
   
The atomic physics aspects have recently been addressed again
\cite{convCMM}.  In that study we pay special attention to
relativistic effects and Coulomb distortion in the wave functions, as
well as parameters of nucleon distributions in various nuclei.

\subsubsection{Experiments}

The experimental searches for the conversion have a long history,
going back to the pioneering cosmic ray experiment by Lagarrigue and
Peyrou \cite{lagarrigue52}.  The progress made in the following 45
years and the future prospects are summarized in  Table
\ref{tab:mueHistory}.

\begin{table}[htbp]
  \caption{History of searches for muon conversion into electrons on
                    nuclei}
  \label{tab:mueHistory}
  \begin{center}
    \leavevmode
    \begin{tabular}[b]{lccrr}
\hline\hline
Year & Muon source & Target  & Upper bound & Ref.  \\ 
\hline
1952 & Cosmic rays & Cu, Sn & $4\times 10^{-2}$ & 
                                        \protect\cite{lagarrigue52}
                 \\[1mm]  
1955 & Nevis cycl. & Cu & $5 \times 10^{-4}$ &
              \protect\cite{steinberger55}
                 \\[1mm]  
1961 & Berkeley synchroc. & Cu & $4\times 10^{-6}$ &
              \protect\cite{sard61}
                 \\[1mm]  
1961-62 & CERN synchroc. & Cu & $2.2\times 10^{-7}$ &
              \protect\cite{conversi61,conforto62}
                 \\[1mm]  
1972 & Virginia SREL synchroc. & Cu & $1.6\times 10^{-8}$ &
              \protect\cite{bryman72}
                 \\[1mm]  
1977 & SIN  & S & $7\times 10^{-11}$ &
              \protect\cite{hahn77}
                 \\[1mm]  
1984 & TRIUMF & Pb & $4.9\times 10^{-10}$ &
              \protect\cite{hargrove84}
                 \\[1mm]  
 &  & Ti & $4.6\times 10^{-12}$ &
                 \\[1mm]  

1992 & PSI  & Pb & $4.6\times 10^{-11}$ &
              \protect\cite{Honecker:1996}
                 \\[1mm] 
1993-97 &  & Ti & $7\times 10^{-13}$ &
              \protect\cite{wintz97}
                 \\[1mm]  
1998-99 (?) & PSI  & Ti, Au (?) & $\sim 10^{-14}$ &
                 \protect\cite{wintz97} 
                 \\[1mm]  
2000 (?) & AGS  & Al & $< 10^{-16}$ & \protect\cite{molzon:97}
                 \\[1mm]  
\hline\hline    \end{tabular}
  \end{center}
\end{table}

The most recent experimental progress has been made in a series of
measurements by the SINDRUM II Collaboration.  The SINDRUM II detector
is characterized by a large solid angle and good momentum resolution
obtained by measuring at least one turn of the helical trajectories of
the decay electrons \cite{Schaff93}.  The limiting factor is the muon
stop rate.  In order to further increase the sensitivity of the
search, a new concept has been developed for the muon source.  This
year the spectrometer is expected to start working with a different
pion beam at PSI.  Muons are obtained from pions decaying in the Pion
Muon Converter (PMC), which is an 8 meter long superconducting
solenoid.  In order to reduce the so called prompt background (high
energy electrons from pion decays) pions are prevented from entering
the spectrometer; this is ensured by a beam stopper at the end of PMC,
where pions which do not decay are absorbed.  These improvements are
expected to permit measurements of $R_{\mu e}$ at the $10^{-14}$
level.

The ideas for a conversion search with sensitivity below $10^{-16}$
were put forward by Lobashev and Dzhilkibaev \cite{Dzhilkibaev:1989}
and resulted in a proposal of an experiment in the Moscow Meson
Factory \cite{Abadzhev:1992}.  This experiment has not been performed
at MMF nor at other places where the proposal was submitted (TRIUMF,
LAMPF, PSI).  Recently it has been realized that the Alternating
Gradient Synchrotron (AGS) facility in Brookhaven may be an ideal
place for this project.  After 1999, AGS becomes an injector for the
Relativistic Heavy Ion Collider (RHIC).  With this task requiring mere
2 hours of AGS time, 22 hours a day will remain for other physics
programs.  Very recently a proposal for an experiment called MECO has
been submitted to BNL \cite{molzon:97}.  It would use the AGS proton
beam for producing high intensity secondary muon beam which would
permit a muon conversion search with a sensitivity better than
$10^{-16}$.

In addition to allowing a search for the muon conversion with more
than 2 orders of magnitude better sensitivity than previous searches,
MECO would serve as a demonstration facility for several elements of
technology needed at a muon collider.  In particular, the muon source
yield of $10^{11}$ muons per second would be about 4 orders of
magnitude more intense that what has been constructed so far.  This
would be a major step towards the $10^{13}\mu/s$ intensity required by
a muon collider.  These intense muon sources require large capture
solenoids; the MECO solenoid would produce a 3 Tesla field.

\section{Anomalous magnetic moment of the muon}

The anomalous magnetic moment of the muon, $a_\mu\equiv (g_\mu-2)/2$,
can in principle be measured directly from the difference of precesion
frequencies of the momentum and spin directions, $\omega_a$.  This
method has been used by three experiments carried out at CERN and is
also being used by the new experiment E821 at Brookhaven.  Results
obtained in past measurements and the accuracy expected to be reached
at E821 are summarized in Table \ref{tab:g2History}.  Since the
magnetic field is calibrated using NMR probes, the actual experiments
measure the ratio of the difference frequency $\omega_a$ and the
proton spin precesion frequency $\omega_p$.  In order to determine
$a_\mu$ the ratio of the proton and muon magnetic moments must also be
measured \cite{Farley92}.

\begin{table}[htbp]
  \caption{Measurements of the muon anomalous magnetic moment}
  \label{tab:g2History}
  \begin{center}
    \leavevmode
    \begin{tabular}[b]{lrr}
\hline\hline
Laboratory & Value & Ref.  \\ 
\hline
 Columbia (Nevis)  & $<5\times 10^{-2}$ & \protect\cite{garwin57}
                 \\[1mm]  
CERN &   $(1162\pm 5)\times 10^{-6}$ & \protect\cite{charpak65}
                 \\[1mm]  
CERN &   $(116616\pm 31)\times 10^{-8}$ & \protect\cite{bailey72}
                 \\[1mm]  
CERN &   $(1165923\pm 8.4)\times 10^{-9}$ & \protect\cite{bailey79,PDG}
                 \\[1mm]  
BNL &   $(?\pm 4-2)\times 10^{-10}$ & \protect\cite{Roberts92,Hughes92}
                 \\[1mm]  
\hline\hline    \end{tabular}
  \end{center}
\end{table}

$a_\mu$ provides both a sensitive test of quantum effects in the
standard model and a window to potential ``new physics'' effects.  The
current experimental value is in good agreement with theoretical
expectations and already constrains physics beyond the standard model
such as supersymmetry and supergravity
\cite{km90,Nath95,Lopez:1994vi,Carena:1997qa,Moroi:1996yh},
dynamical or loop muon mass generation \cite{MassMech}, compositeness
\cite{Brodsky:1980zm,Gonzalez-Garcia:1996rx}, leptoquarks
\cite{Couture95}, two-Higgs-doublet extensions of the standard model
\cite{Krawczyk:1997sm} etc.

The experiment E821 at Brookhaven National Laboratory which has
recently began is expected to reduce the uncertainty in $a_\mu^{\rm
exp}$ to roughly $\pm 40\times 10^{-11}$, with one month of dedicated
running.  With subsequent longer dedicated runs it could statistically
approach the anticipated systematic uncertainty of about $\pm
10-20\times 10^{-11}$ \cite{Bunce}.  At those levels, both electroweak
one and two loop effects become important and ``new physics'' at the
multi-TeV scale is probed.  Indeed, generic muon mass generating
mechanisms (via perturbative or dynamical loops \cite{MassMech}) lead
to $\Delta a_\mu \approx m_\mu^2/\Lambda^2$, where $\Lambda$ is the
scale of ``new physics''.  At $\pm 40\times 10^{-11}$ sensitivity,
$\Lambda \approx 5$ TeV is being explored.

To fully exploit the anticipated experimental improvement, the
standard model prediction for $a_\mu$ must be known with comparable
precision.  That requires detailed studies of very high order QED
loops, hadronic effects, and electroweak contributions through two
loop order.  The contributions to $a_\mu$ are traditionally divided
into
\begin{eqnarray}
a_\mu=a_\mu^{\rm QED}+a_\mu^{\rm Hadronic}+a_\mu^{\rm EW}
\end{eqnarray}
QED loops have been computed analytically to the sixth order
and to an even higher order numerically
\cite{Kinoshita:1990wp,Kinoshita:1993pq,Samuel:1991qf,%
Laporta:1993pa,Laporta:1993,Kinoshita95,Laporta:1996mq}
\begin{eqnarray}
a_\mu^{\rm QED} &=& {\alpha\over 2 \pi}
+0.765857381(51) \left( {\alpha\over  \pi}\right)^2
+24.050531(40) \left( {\alpha\over  \pi}\right)^3
\nonumber\\ &&
+ 126.02(42) \left( {\alpha\over  \pi}\right)^4
+ 930(170) \left( {\alpha\over  \pi}\right)^5
\label{eq:qed}
\end{eqnarray}
Employing $\alpha= 1/137.03599944(57)$ obtained from the electron
$g_e-2$, implies \cite{Kinoshita95}
\begin{eqnarray}
a_\mu^{\rm QED} &=& 116584706(2)\times 10^{-11}
\end{eqnarray}
The uncertainty is well within the $\pm 20-40\times 10^{-11}$ goal.

Hadronic vacuum polarization corrections to $a_\mu$ enter at ${\cal
O}(\alpha/\pi)^2$.  They can be evaluated via a dispersion relation
using $e^+e^- \to hadrons$ data and perturbative QCD (for the very
high energy regime).  Recent analysis of $e^+e^-$ data \cite{Jeg95}
and hadronic $\tau$ decays \cite{Alemany:1997tn} (including hadronic
corrections in two-loop QED diagrams \cite{kinoshita85,Krause:1997rf})
gives
\begin{eqnarray}
a_\mu^{\rm Hadronic}({\rm vac.\,pol.})= 6911(100)\times 10^{-11}
\label{eq:hadr}
\end{eqnarray}
The accuracy of the estimate of the hadronic contribution has not yet
reached the desired level.  Ongoing improvements in $e^+e^- \to
hadrons$ measurements at low energies along with additional
theoretical input should significantly lower the uncertainty in
(\ref{eq:hadr}).

The result in (\ref{eq:hadr}) must be supplemented by hadronic light
by light amplitudes (which are of three loop origin) \cite{light}.
The recent estimates \cite{bijn95,Hayakawa:1997rq} agree with each
other within error bars; here we employ the value obtained in
\cite{Hayakawa:1997rq}
\begin{eqnarray}
a_\mu^{\rm Hadronic}({\rm light\, by \, light})= -79(15)\times
10^{-11}
\label{eq:light}
\end{eqnarray}
Combining (\ref{eq:hadr}) and (\ref{eq:light}) leads to the total
hadronic contribution
\begin{eqnarray}
a_\mu^{\rm Hadronic}= 6832(101)\times 10^{-11}
\end{eqnarray}

The main objective of the new experiment at BNL is to examine the
electroweak contributions to $a_\mu$.
At the one loop level, the standard model predicts 
\cite{fls72,Jackiw72,ACM72,Bars72,Bardeen72}
\begin{eqnarray}
a_\mu^{\rm EW}(\rm 1\,loop) &=&
{5\over 3}{G_\mu m_\mu^2\over 8\sqrt{2}\pi^2}
\left[1+{1\over 5}(1-4s_W^2)^2
+ {\cal O}\left({m_\mu^2 \over M_{W,H}^2}\right) \right]
\nonumber \\
& \approx & 195 \times 10^{-11}
\label{eq:oneloop}
\end{eqnarray}
where $G_\mu = 1.16639(1) \times 10^{-5}$ GeV$^{-2}$, 
and the weak mixing angle
$\sin^2\theta_W\equiv s_W^2 = 1-M_W^2/M_Z^2=0.224$.  We can safely
neglect the ${\cal O}\left({m_\mu^2 / M_{W,H}^2}\right)$ terms in
(\ref{eq:oneloop}).

The one loop result in (\ref{eq:oneloop}) is about five to ten times
the anticipated experimental error.  Naively, one might expect higher
order (2 loop) electroweak contributions to be of relative ${\cal
O}(\alpha/ \pi)$ and hence negligible; however, that is not the case.
Kukhto, Kuraev, Schiller, and Silagadze (KKSS) \cite{KKSS} have shown
that some two loop electroweak contributions can be quite large and
must be included in any serious theoretical estimate of $a_\mu^{\rm
EW}$ or future confrontation with experiment.

The two loop electroweak contributions to $a_\mu^{\rm EW}$ naturally
divide into so-called fermion and boson parts
\begin{eqnarray}
a_\mu^{\rm EW} = a_\mu^{\rm EW} ({\rm 1\,loop})
+a_\mu^{\rm EW}({\rm 2\,loop;\,ferm.})
+a_\mu^{\rm EW}({\rm 2\,loop;\,bos.})
\end{eqnarray}
The $a_\mu^{\rm EW}({\rm 2\,loop;\,ferm.})$ includes all two loop
electroweak corrections which contain closed fermion loops while all
other contributions are lumped into $a_\mu^{\rm EW}({\rm
2\,loop;\,bos.})$.  The fermio\-nic correction $a_\mu^{\rm EW}({\rm
2\,loop;\,ferm.})$ was calculated in \cite{CKM95,Peris:1995bb}.  For
$M_{\rm Higgs} \approx$ 250 GeV it reduces $a_\mu^{\rm EW}$ by 11.8\%.
More recently that effort has been completed by computing the bosonic
contribution, $a_\mu^{\rm
EW}({\rm 2\,loop;\,bos.})$ \cite{CKM96}.  Combining these results
leads to a total reduction of $a_\mu^{\rm EW}$ by a factor
$(1-97\alpha/\pi) \approx 0.77$ and the new electroweak prediction
\begin{eqnarray}
a_\mu^{\rm EW} = 151(4) \times 10^{-11}
\label{eq:newaEW}
\end{eqnarray}
The assigned error of $\pm 4\times 10^{-11}$ is due to uncertainties
in $M_H$ and quark two loop effects.  It also allows for possible
three loop (or higher) electroweak contributions. 

The present theoretical prediction for the muon anomalous magnetic
moment is  
\begin{eqnarray}     
a_\mu^{\rm theory} = 116591689(101) \times 10^{-11}
\end{eqnarray}
with extremely small QED and EW uncertainties.  What remains is to
reduce the hadronic uncertainty by a factor of 3 (or more) via
improved $e^+e^-\to hadrons$ data and additional theoretical input.
Then, one can fully exploit the anticipated improvement in
$a_\mu^{\rm exp}$ from E821 at Brookhaven.

\section{Conclusions}
I have reviewed two aspects of muon physics: searches for muon number
non-conserva\-tion and measurements of its anomalous magnetic moment.
The new measurement of $g-2$ at Brookhaven at the level of accuracy
improved by a factor of 20 or more will test the quantum corrections
in the standard model and impose bounds on possible new physics.  New
ideas for background suppression in $\mu\to e\gamma$ searches might
permit a stringent test of supersymmetric grand unification models.
Most exciting is the proposal for searching for muon conversion into
electrons in the nuclear field at the level of $10^{-16}$ or lower.
Such an experiment would test many proposed extensions of the standard
model at an unprecedented level of precision.  With these new ideas
muon research remains one of the most interesting areas in particle
physics.

\section*{Acknowledgements}
I am grateful to W. Marciano, K. Melnikov, and B. Krause for
discussions and collaboration on various aspects of muon physics, and
to T.~Koz{\l}owski for explaining some details of the SINDRUM II
spectrometer.  I wish to thank W. Marciano and P. Osland for making my
participation in the conference ``Beyond the Standard Model V''
possible.  This research has been supported by BMBF grant
BMBF-057KA92P and by DOE grant DE-AC02-76CH00016.

\end{document}